\documentclass[useAMS,usenatbib,onecolumn]{mn2e}
\usepackage{psfig}
\usepackage{epsfig}

\title[21cm Angular Power Spectrum]
  {Contribution of Cross-Correlations to the 21cm Angular Power 
Spectrum in the Epoch of Reionization}

\author[Zheng]
  {Qian Zheng$^{1,2}$
\\
  $^1$National Astronomical Observatories, Chinese Academy of Sciences, Beijing 
100012, China\\
  $^2$Graduate School of Chinese Academy of Sciences, Beijing 100049, China
 }

\begin{document}

\date{Received date / accepted date}

\pagerange{\pageref{firstpage}--\pageref{lastpage}} \pubyear{2009}

\label{firstpage}

\maketitle

\begin{abstract}
Measurement of the 21cm hyperfine transition of neutral hydrogen provides 
a unique probe of the epoch of reionization and the Dark Ages.
Three major mechanisms are believed to dominate the radiation process:
emission from neutral hydrogen surrounding the ionized 
bubbles of first galaxies and/or quasars, emission from
neutral hydrogen inside minihalos, and absorption of diffuse 
neutral hydrogen against the cosmic microwave background. 
In the present work, by simply combining the existing analytic models for
the three mechanisms, we investigate the contribution of 
cross-correlation between these three components to the total
21cm angular power spectrum, in the sense that neutral hydrogen
associated with different radiation processes traces the 
large-scale structures of underlying density perturbations. 
While the overall 21cm power spectrum remains almost unchanged 
with the inclusion of the cross-correlations, the cross-correlation
may play a key role in the determination of the 21cm power
spectrum during the transition of 21cm radiation 
from emission-dominated phase to absorption-dominated phase
at redshift $z\approx 20$. A significant suppression 
in the 21cm angular power spectrum during this transition is
anticipated as the result of negative contribution of 
the cross-correlation between the absorption of diffuse neutral 
hydrogen and the emission components. Therefore, an accurate 
prediction of the cosmic 21cm power spectrum  should take
the cross-correlation into account especially at the transition
phase. 
\end{abstract}


\section{Introduction}

The 21cm hyperfine transition of neutral hydrogen 
in the form of either emission or absorption against the cosmic
microwave background (CMB) provides a unique 
probe of the history of our universe between the surface of last 
scattering ($z\approx1000$) and the end of reionization ($z\approx6$).
While the expected brightness temperature of these 21cm signals 
is two orders of magnitude smaller than the CMB temperature and 
can be easily swamped by extremely strong foreground such as 
the Milky Way, many ambitious radio facilities 
(e.g. 21CMA\footnote{See http://cosmo.bao.ac.cn}, 
LOFAR\footnote{See http://www.lofar.org},
MWA\footnote{See http://www.haystack.mit.edu/arrays/MWA},
PAPER\footnote{See http://astro.berkeley.edu/$\sim$dbacker/eor}, 
SKA\footnote{See http://www.skatelescope.org}, etc) have been constructed 
and are being planned to 
search for the weak signals primarily from the epoch of reionization
(EOR) at $6<z<20$. Indeed, the Dark Ages and the EOR
may constitute the last frontier of observational cosmology. 

There have been many efforts - both theoretical and numerical - 
over the past few years aimed at understanding 
various physical processes in the intergalactic medium (IGM) 
during the EOR [see \citealt{furlanetto06} 
and \citealt{morales} for a review].  
In order to make the 21cm signals visible,  
the spin temperature of neutral hydrogen must differ from the 
temperature of the CMB, $T_{\rm CMB}$, and an effective mechanism  
exists to couple the spin temperature to the gas temperature. 
Apart from some exotic scenarios such as
dark matter particle decay, three mechanisms are believed to 
play dominant roles in the generation of observable 21cm signals 
from the Dark Ages to the EOR: scattering of UV photons of 
first stars primarily through the so-called Wouthuysen-Field effect 
(\citealt{wouthuysen};  \citealt{field}), 
heating of IGM by collision 
mainly within minihalos (\citealt{madau}; 
\citealt{iliev02}), and absorption of uncollapsed IGM 
against the CMB radiation prior to the formation of 
first cosmic structures (\citealt{scott}; \citealt{loeb}; Madau et al. 1997; 
\citealt{lewis}). 
Among these three mechanisms, the foreground absorption of 
neutral hydrogen against the CMB dominates the 21cm background
through the epoch of the Dark Ages down to redshift $z\approx20$, 
while for $6<z<20$ the energy release of first luminous objects 
(stars and/or quasars) plays a leading role in generating the 21cm 
emission in excess of the CMB (e.g. \citealt{scott}; 
\citealt{barkana}). During the transition phase or 
the 'Grey Ages' between the end of the Dark Ages and the dawn of 
reionization at $z\approx20$, a combined mechanism may come into
effect in the redshifted 21cm background.

If we decompose the surface brightness temperature of 21cm background 
from the Dark Ages through the EOR into three components:
emission from neutral hydrogen surrounding the ionized 
bubbles of first galaxies and/or quasars $T_{\rm em}$, emission 
inside minihalos  $T_{\rm mh}$, and absorption of IGM against
the CMB  $T_{\rm ab}$,  the 21cm power spectrum can be formally determined by 
three auto-correlation terms $\langle T_{\rm em}T_{\rm em}\rangle+
\langle T_{\rm mh}T_{\rm mh}\rangle+\langle T_{\rm ab}T_{\rm ab}\rangle$,
and three cross-correlation terms 
$2\langle T_{\rm em}T_{\rm mh}\rangle+
2\langle T_{\rm em}T_{\rm ab}\rangle+2\langle T_{\rm mh}T_{\rm ab}\rangle$.
In particular, two of the cross-correlations relevant to the absorption
have negative amplitudes as a result of 
$T_{\rm S}<T_{\rm CMB}$, where $T_{\rm S}$ is the spin temperature of 
neutral hydrogen. Although quite a lot of work has been done
in recent literature on the theoretical estimate of the three
auto-correlation terms (e.g. \citealt{iliev02}; 
\citealt{zaldarriaga}; 
\citealt{loeb}; \citealt{furlanetto04a, furlanetto04b}; 
\citealt{furlanetto06};
\citealt{santos}; \citealt{lewis}; \citealt{mao}), 
it has remained unclear so far to what extent 
the three cross-correlations contribute to the total 21cm power 
spectrum, especially in the Grey Ages. 
The essentials of three cross-correlations arise from the fact that
both collapsed and uncollapsed neutral hydrogen may trace underlying 
gravitational potentials of large-scale density perturbations, 
through which their 21cm emission/absorption signals are
correlated at large-scales.
\cite{furlanetto06} have actually included
the cross-correlation between $T_{\rm em}$ and $T_{\rm mh}$ 
in the study of the power spectrum of 21cm fluctuations,  
although their work focused  mainly on the observational distinction between 
signatures of minihalos and those of IGM, in which massive halos (galaxies)
are ignored. 
In the present study we will concentrate on the evaluation of 
each of the cross-correlations in terms of the 21cm angular power 
spectrum for different frequencies or redshifts.

Note, however, that two of the cross-correlations, 
$\langle T_{\rm em}T_{\rm ab}\rangle$ and 
 $\langle T_{\rm em}T_{\rm mh}\rangle$, 
should be safely treated as a mathematical elegance 
rather than a physically realistic situation through most of 
the period of the Dark Ages to the EOR. The latter requires  
the co-existence of 21cm emission from 'hot' IGM around 
first galaxies and 21cm absorption from 'cold' IGM, represented by
$\langle T_{\rm em}T_{\rm ab}\rangle$, and
the co-existence of 21cm emission from 'hot' IGM around 
first galaxies and 21cm emission from 'hot' IGM inside minihalos, 
represented by  $\langle T_{\rm em}T_{\rm mh}\rangle$, within the
same regions. Unless an inhomogeneous heating scenario is considered 
(\citealt{barkana}; \citealt{pritchard07};
\citealt{santos}; \citealt{mesinger}), 
the two correlation terms $\langle T_{\rm em}T_{\rm ab}\rangle$ and  
$\langle T_{\rm em}T_{\rm mh}\rangle$ may become physically meaningless 
except for the very rapid transition phase or the Grey Ages. 
In this regard, the last cross-correlation between the 21cm
emission from minihalos and the 21cm absorption from the diffuse
IGM,  $\langle T_{\rm mh}T_{\rm ab}\rangle$, is probably the only physically
plausible term that is worth investigating during and even 
before the Grey Ages.

The main difficulty of evaluating accurately the cross-correlations 
at present within the framework of analytic description of the 
reionization process 
is perhaps the absence of a unification scheme in which the
three mechanisms of generating cosmic 21cm signals can be incorporated
and treated at any redshifts through the Dark Ages to the EOR. 
The emission scenario associated with first galaxies which has been
extensively discussed in recent literature applies only to lower
redshifts at $6<z<20$, while the absorption model of uncollapsed 
IGM is restricted to higher redshifts beyond $z\approx20$. During the 
Grey Ages when the effect of the cross-correlations between different 
mechanisms is expected to be visible in the cosmic 21cm power spectrum, 
our theoretical prediction relies on the extrapolation of 
the current emission and absorption models unless we make an attempt      
at exploring the unification scheme in which the distribution and 
evolution of both neutral hydrogen and its reionization can be correctly 
described through the entire period of the Dark Ages and the EOR including 
the transition phase.  It is likely that numerical simulation 
(e.g. \citealt{zahn}; \citealt{trac}; 
\citealt{iliev08}; \citealt{santos}) should
be eventually used so that various physical processes can be 
incorporated in the estimate of the cross-correlations in 
the cosmic 21cm power spectrum. 

Having realized these difficulties and limitations,  we still 
intend to take the three cross-correlations into account 
in the calculation of the cosmic 21cm power spectrum. 
By simple combining the existing analytic scenarios of generating the 
cosmic 21cm signals at the Dark Ages and the EOR,
we address the question of what and how large observational effect 
one may expect to see if the three correlation terms are included.
Our goal in the present paper is to call the attention of 
the 21cm cosmology community to the possible role of 
the cross-correlations in the 21cm power spectrum especially 
at the Grey Ages rather than the detailed modelling of 
the evolution of cosmic neutral hydrogen and reionization process. 
It is hoped that future numerical simulation can provide a
more sophisticated treatment of the problem. 

%%%%%%%%%%%%%%%%%%%%%%%%%%%%%%%%%%%%%%%%%%%%%%%%%%%%%%%%%%%%%%%%%

\section{General Framework}

We concentrate our attention on the following three mechanisms only:
21cm emission from the neutral hydrogen  surrounding the Str\"omgren 
spheres of the first galaxies/quasars, 21cm emission from minihalos, 
and 21cm absorption against the CMB from the uncollapsed IGM. If we use
the surface brightness temperature to denote the cosmic 21cm radiation,
the total background can be written as 
%1  
\begin{equation}
T=T_{\rm em}+T_{\rm mh}+T_{\rm ab}.
\end{equation}
We usually work with the two-point correlation function
or power spectrum in the Fourier space to extract the
considerably faint 21cm signals in a statistical manner. This yields
%2
\begin{equation}
\langle TT\rangle=  \langle T_{\rm em}T_{\rm em}\rangle+
                    \langle T_{\rm mh}T_{\rm mh}\rangle+
                    \langle T_{\rm ab}T_{\rm ab}\rangle+
                    2\langle T_{\rm em}T_{\rm mh}\rangle+
                   2\langle T_{\rm em}T_{\rm ab}\rangle+
                   2\langle T_{\rm mh}T_{\rm ab}\rangle.
\end{equation}
or in terms of power spectrum 
%3 
\begin{equation}
P(\mbox{\boldmath $k$})= P_{\rm em-em}(\mbox{\boldmath $k$})+
                         P_{\rm mh-mh}(\mbox{\boldmath $k$})+
                         P_{\rm ab-ab}(\mbox{\boldmath $k$})+
                        2P_{\rm em-mh}(\mbox{\boldmath $k$})+
                        2P_{\rm em-ab}(\mbox{\boldmath $k$})+
                        2P_{\rm mh-ab}(\mbox{\boldmath $k$}).
\end{equation}

In what follows we utilize the angular power spectrum $C(\ell)$ by 
specifying the observing frequency $\nu$ (or redshift)  instead of 
the 3D power spectrum $P(\mbox{\boldmath $k$})$. Moreover, we adopt a 
small-angle approximation, which allows us
to convert the 3D power spectrum $P(\mbox{\boldmath $k$})$ 
of a perturbation field 
into the corresponding angular power spectrum $C(\ell)$ in terms of the 
Limber's equation. If the frequency response function of 21cm experiment 
is $W_{\nu_0}(\nu)$ or $W_{r_{\rm 0}}(r)$ centered on $\nu_{\rm 0}$ or $r_{\rm 0}$,
where $r$ is the comoving distance, then two-point angular power
spectrum can be obtained by
%4
\begin{equation}
C(\ell)= \int dr \frac{W^2_{r_{\rm 0}}(r)}{r^2}P({\ell}/r).
\end{equation}
In fact, in the calculation of both auto and cross angular power 
spectra of the 21cm background, many of the perturbations can be 
related to the underlying matter density fluctuations. 
Consequently, in most cases $P(\mbox{\boldmath $k$})$ simply represents 
the power spectrum of 
dark matter, and other coefficients can be absorbed into the window function.
For the window function, we will take a Gaussian form of
%5
\begin{equation}
W_{r_{\rm 0}}(r)=\frac{1}{\sqrt{2\pi}\Delta r}e^{-(r-r_{\rm 0})^2/2\Delta r^2},
\end{equation}
in which the resolution $\Delta r$ in $r$ is related to the 
observing frequency bandwidth $\Delta \nu$ through
%6
\begin{equation}
\Delta r= 1.725\left(\frac{\Delta\nu}{0.1{\rm MHz}}\right)
          \left(\frac{1+z}{10}\right)^{1/2}
          \left(\frac{\Omega_{\rm M}h^2}{0.15}\right)^{1/2}\,{\rm Mpc}.
\end{equation}

We evaluate the power spectrum of density perturbation following
the analytical halo approach to include non-linear structures
(\citealt{seljak}; \citealt{cooray}).
The non-linear dark matter power spectrum $P_{\rm M}({\mbox{\boldmath $k$}},z)$ 
consists of a single halo term $P_{\rm M}^{\rm 1h}({\mbox{\boldmath $k$}},z)$ 
plus a clustering term $P_{\rm M}^{\rm 2h}({\mbox{\boldmath $k$}},z)$:
%7 8
\begin{eqnarray}
P_{\rm M}^{\rm 1h}({\mbox{\boldmath $k$}},z)&=&\int dM \frac{d^2n(M,z)}{dMdV} 
                  \left|\frac{\rho_{\rm h}(M,z,{\mbox{\boldmath $k$}})}
                             {\overline{\rho}(z)}\right|^2 \\
P_{\rm M}^{\rm 2h}({\mbox{\boldmath $k$}},z)&=&P_{\rm M}^{\rm lin}
                  ({\mbox{\boldmath $k$}},z)  
                   \left[\int dM \frac{d^2n(M,z)}{dMdV} b(M,z)
                     \frac{\rho_{\rm h}(M,z,{\mbox{\boldmath $k$}})}
                    {\overline{\rho}(z)}\right]^2
\end{eqnarray}
where $\rho_{\rm h}(M,z,{\mbox{\boldmath $k$}})$ is the Fourier transform 
of dark matter density
profile $\rho_{\rm h}(M,z)$ of a halo $M$ at $z$, and $\overline{\rho}(z)$
is the mean mass density of the background universe at $z$.  
We adopt the functional form suggested by
numerical simulations (\citealt{navarro})
for $\rho_{\rm h}(M,z)$ and fix the concentration parameter in terms of 
the empirical fitting formula of  \cite{bullock}. 
$b(M,z_{\rm s})$ is the bias parameter, for which we use
the analytic prescription of \cite{mo}. 
The mass function of dark halos $d^2n(M,z)/dMdV$ is assumed to 
follow the standard Press-Schechter function
%9
\begin{equation}
\frac{d^2n(M,z)}{dMdV}=-\sqrt{\frac{2}{\pi}} \frac{\bar{\rho}}{M} 
    \frac{\delta_{\rm c}}{\sigma^2} 
    \frac{d\sigma}{dM} 
    \exp{\left(-\frac{\delta_c^2}{2\sigma^2} \right)},
\end{equation}
in which $\sigma$ is the linear theory variance of the mass density
fluctuation in sphere of mass $M$:
%10
\begin{equation}
\sigma^{2}=
(1/2\pi^2) \int_0^{\infty}k^2 P_{\rm M}^{\rm lin}(k){\vert W_R(kR)\vert}^2 dk,
\end{equation}
and $W_R(kR)$ is the Fourier representation of the window function. For
a top-hat window function, we have
$W_{\rm R}(kR)=3[\sin(kR)-kR\cos (kR)]/(kR)^3$. We use the present-day linear 
matter power spectrum $P_{\rm M}^{\rm lin}(k)$ given by \cite{bardeen} 
and a power-law form ($\propto k^{n_{\rm s}}$) for the primordial matter density 
fluctuation.  Moreover, we adopt the concordance cosmological
model ($\Lambda$CDM) with the following choice of cosmic parameters: 
$\Omega_{\rm M}=0.27$, $\Omega_{\Lambda}=0.73$, 
$\Omega_bh^2=0.0224$, $h=0.71$, $n_s=1.0$ and $\sigma_8=0.84$.

Since our attention is concentrated  on the evaluation of
the 21cm angular power spectrum of each of the cross-correlations rather
than the detailed modelling of the underlying physical process
of 21cm transition, we adopt the existing analytic algorithms developed
in the recent literature to calculate the cosmic 21cm emission/absorption
signal. In what follows we summarize very briefly each of the
analytic or semi-analytic approaches in the computation of 
auto-correlation angular power spectrum and extend the work to 
the evaluation of the corresponding cross-correlations.

%%%%%%%%%%%%%%%%%%%%%%%%%%%%%%%%%%%%%%%%%%%%%%%%%%%%%%%%%%%%%%%%%
\section{Power Spectra}

\subsection{Auto-correlation: $C_{\rm em-em}(\ell)$}

We follow essentially the prescription of Zaldarriaga et al.(2004) to
estimate the 21cm emission angular power spectrum from neutral hydrogen
surrounding the ionized bubbles of first galaxies. In their treatment, 
the reionization process is associated with the growth of HII regions 
around individual galaxies, manifested by a sphere of radius
of $R$. We use an analytic form of 
$R=\{3+0.424[1-\exp[(\bar{x}_{H}-0.5)/0.375]^2]\}$ $h^{-1}$Mpc 
to account for the cosmic evolution of $R$, which roughly matches
the values adopted by  Zaldarriaga et al. (2004) 
at different redshifts.
The overall cosmic evolution of reionization is specified by
the mean neutral hydrogen fraction, for which a simple analytic 
expression is assumed:
$\bar{x}_{\rm H}(z)=1/{1+exp[-(z-z_{\rm 0})/\Delta{z}]}$, with a choice of 
$z_0=10$ and $\Delta{z}=0.5$ in our calculation.  The brightness
temperature in excess of $T_{\rm CMB}$ is 
% 11 12
\begin{eqnarray}
T_{\rm em}&=&T_{\rm 0}(1+\delta)x_{\rm H}\\
T_0&=&{23}{\rm mK}
      \left(\frac{\Omega_{\rm b}h^{2}}{0.02}\right)
      \left(\frac{0.15}{\Omega_{\rm M}h^{2}}\right)^{1/2}
      \left(\frac{1+z}{10}\right)^{1/2},
\end{eqnarray}
in which we have assumed that the spin temperature of neutral hydrogen is
much higher than $T_{\rm CMB}$, and the peculiar motion of the IGM is neglected.
The perturbation in $T_{\rm em}$ arises from 
a combination of fluctuations in both the matter density $(1+\delta)$ 
and the neutral fraction $x_{\rm H}$, giving rise to the power spectrum 
%13
\begin{equation}
P_{(1+\delta)x_{\rm H}}(\mbox{\boldmath $k$})
      =\bar{x}_{\rm H}^{2}P_{\rm M}(\mbox{\boldmath $k$})+
      (\bar{x}_{\rm H}-\bar{x}_{\rm H}^{2})P_{x_{\rm H}\delta}(\mbox{\boldmath $k$})+
      (\bar{x}_{\rm H}-\bar{x}_{\rm H}^{2})P_{x_{\rm H}}(\mbox{\boldmath $k$}).
\end{equation}
Namely, the power spectrum of $(1+\delta)x_{\rm H}$ comprises two 
auto-correlations and one cross-correlation between the two fields.
An analytic model was used to construct 
$P_{{x_{\rm H}}}(\mbox{\boldmath $k$})$ and 
$P_{x_{\rm H}\delta}(\mbox{\boldmath $k$})$ in Zaldarriaga et al. (2004)
and a refined model was proposed in their subsequent work 
(Furlanetto et al. 2004a).  
The observed 21cm brightness temperature in the 
direction $\mbox{\boldmath $\theta$}$ at frequency $\nu$ is thus
% 14
\begin{equation}
T^{\rm obs}_{\rm em}(\mbox{\boldmath $\theta$},\nu)= 
      \int dr W_{\nu}(r)T_{\rm em}(\mbox{\boldmath $\theta$},\nu).
\end{equation}
Applying the Limber's equation, 
we can obtain the corresponding angular power spectrum $C_{\rm em-em}(\ell)$
by replacing $P({\ell}/r)$ with ${T_0}^{2}P_{(1+\delta)x_{\rm H}}({\ell}/r)$ 
in Eq.(4).

\subsection{Auto-correlation: $C_{\rm mh-mh}(\ell)$}

It is argued that collisional excitation of the neutral hydrogen 
inside minihalos may constitute a non-negligible fraction of the cosmic 21cm
background, especially at the Grey Ages. Many investigations, 
based on both semi-analytical treatment and numerical simulations,
have been made towards the modeling of the radiation process and
the signatures of the resultant 21cm background (e.g. 
\citealt{carilli}; \citealt{iliev02,iliev03}; 
\citealt{ciardi}; \citealt{furlanetto06}; \citealt{shapiro}) 
Minihalos are restricted by two mass thresholds: 
the Jeans mass that sets the low-mass limit on collapsed halos, 
and the upper limit beyond which the atomic cooling becomes 
efficient. In the $\Lambda$CDM cosmological model, the two 
mass limits can be estimated by \cite{shapiro}
%15 16
\begin{eqnarray}
M_{\rm min}&=&5.7\times{10^{3}}
         \left(\frac{\Omega_{\rm M}h^{2}}{0.15}\right)^{-1/2}
         \left(\frac{\Omega_{\rm b}h^{2}}{0.02}\right)^{-3/5}
         \left(\frac{1+z}{10}\right)^{3/2}\;M_{\odot},\\
M_{\rm max}&=&3.9\times{10^{7}}
         \left(\frac{\Omega_{\rm M}h^{2}}{0.15}\right)^{-1/2}
         \left(\frac{1+z}{10}\right)^{-3/2}M_{\odot}.
\end{eqnarray}

It is unlikely that present and even future radio telescopes/arrays 
will be able to resolve minihalos because of their small size 
and low surface brightness ($<\mu$Jy). We therefore deal with the 
total 21cm flux of each minihalo, and neglect the contribution of 
the Poisson term in the calculation of power spectrum. 
In order to simplify the procedure, also guided by the result of 
numerical simulations, we assume that the 21cm 
flux $F$ of a minihalo is proportional to its total mass $M$. We then 
fit the simulated line-integrated flux $F$
versus minihalo mass $M$ of \cite{iliev02} to the following
linear relation in the Log F-Log M space:
%17
\begin{equation}
\log F=\log M + E(z),
\end{equation}
where the only free parameter $E(z)$, which characterizes the cosmic
evolution of minihalos, is determined and extrapolated at each redshift 
in terms of Figure 1 of \cite{iliev02}.
Note that the cosmological parameters adopted in \cite{iliev02}
are slightly different from ours. Here we have neglected the 
possible modification to the above relationship. 
The overall surface brightness temperature of all minihalos is given by
%18
\begin{equation}
T_{\rm mh}=  \left(\frac{\lambda^2}{2k_B}\right)
           \frac{\lambda(1+z)^2}{c}
           \frac{dV}{dzd\Omega}
            \int_{\rm Mmin}^{\rm Mmax}F\left[\frac{d^2n(M,z)}{dMdV}\right]dM,
\end{equation}
where $k_{\rm B}$ is the Boltzmann constant, and $dV/dzd\Omega$ is the
comoving volume per unit redshift per unit solid angle. 
Note that we have converted the total flux into the surface 
brightness temperature in terms of the Rayleigh-Jeans approximation. 
The angular power spectrum of $T_{\rm mh}$ can be obtained 
using the Limber's equation
%19
\begin{equation}
C_{\rm mh-mh}(\ell)= \left(\frac{\lambda^3}{2k_B}\right)^2
          \int  \frac{(1+z)^4}{H^2(z)}
             W_{\nu}^{2}(r)f^2(r) P_{\rm M}(\ell/r) r^2 dr,
\end{equation}
where $f(r)\equiv\int_{\rm Mmin}^{\rm Mmax}F[\frac{d^2n(M,z)}{dMdV}]b(M,z)dM$, 
and $H(z)$ is the Hubble constant at $z$.

\subsection{Auto-correlation: $C_{\rm ab-ab}(\ell)$}

It was formulated long time ago that before the formation of first
stars at $z\approx20$, neutral hydrogen may absorb the CMB flux 
since the baryon temperature dropped faster than the CMB temperature
in the framework of adiabatic cooling of baryon with 
evolution of the universe (\citealt{field}). The fluctuation of 
such a 21cm absorption field arises purely from the underlying
matter density perturbation, $T_{\rm ab}=(dT_{\rm ab}/d\delta)\delta$.
The 21cm absorption surface brightness temperature along the 
line-of-sight is 
%20
\begin{equation}
T_{\rm ab}(\mbox{\boldmath $\theta$},\nu)=
      \int dr W_{\nu}(r)\frac{dT_{\rm ab}}{d\delta}
      \delta(\mbox{\boldmath $\theta$},\nu).
\end{equation}
The corresponding angular power spectrum can be derived using
the Limber's equation
%21
\begin{equation}
C_{\rm ab-ab}(\ell)= \int\frac{W_{\nu}^{2}(r)}{r^{2}}
       \left(\frac{dT_{\rm ab}}{d\delta}\right)^{2}
        P_{\rm M}({\ell}/r)dr.
\end{equation}

Finally we can use essentially the approach of 
\cite{loeb} to calculate the brightness 
temperature variation with respect to $\delta$, $dT_{\rm ab}/d\delta$.

\subsection{Cross-correlation: $C_{\rm em-mh}(\ell)$}

Neutral hydrogen both surrounding the ionized bubbles of first 
galaxies/quasars and inside minihalos contributes to the  
cosmic 21cm emission. If galaxies/quasars (massive halos) 
and minihalos trace common gravitational potentials of 
underlying density perturbations at large scales, their 21cm 
emission signals should be strongly correlated, and the 
combined effect raises the amplitude of the total 21cm power spectrum. 
Yet,  the above scenario is too simplistic because it
is implicitly assumed that the minihalos may survive within HII
regions for relatively long time. This may happen 
during the Grey Ages when the population 
of first luminous galaxies are rare and evaporation process is 
still inefficient. Another possibility is that minihalos were not
yet swallowed by the expanding bubbles of first galaxies/quasars 
and both of them shared the same density perturbation field.
The cross-correlation $C_{\rm em-mh}(\ell)$
eventually vanishes with the evaporation process of minihalos. 
A combination of Eqs.(11), (12) and (18) yields
%22
\begin{equation}
\langle T_{\rm em}T_{\rm mh}\rangle= T_0(1+\delta)x_{\rm H}
        \left(\frac{\lambda^3}{2k_B}\right)
        \frac{(1+z)^2r^2}{H(z)}
         \int_{\rm Mmin}^{\rm Mmax}F\left[
         \frac{d^2n(M,z)}{dMdV}\right]dM 
\end{equation}
Replacing $\frac{d^2n(M,z)}{dMdV} $by $\frac{d^2\bar{n}}{dMdV}[1+b(M,z)\delta]$ 
and removing the global component, we can easily derive the 3D power spectrum
%23
\begin{equation}
P_{\rm em-mh}({\mbox{\boldmath $k$}})= T_{\rm 0}\left(\frac{\lambda^3}{2k_{\rm B}}
        \right)
        \frac{(1+z)^2r^2}{H(z)}
        [(f+g)P_{x_{\rm H}\delta}({\mbox{\boldmath $k$}})+
        \bar{x}_{\rm H}f P_{\rm M}
        ({\mbox{\boldmath $k$}})],
\end{equation}
in which $g\equiv\int_{\rm Mmin}^{\rm Mmax}F[\frac{d^2{\bar{n}}(M,z)}{dMdV}]dM$. 
Furthermore, the corresponding angular power spectrum under the 
small-angle approximation is
%24
\begin{equation}
C_{\rm em-mh}(\ell)= T_0\left(\frac{\lambda^3}{2k_B}\right)
        \int \frac{(1+z)^2W_{\nu}^{2}(r)}{H(z)}
        [(f+g)P_{{x}_{\rm H}\delta}(\ell/r)
        +\bar{x}_{\rm H}f P_{\rm M}(\ell/r)]dr.
\end{equation}
Note again, however, that the above expressions apply only to the regimes 
outsides the ionized bubbles surrounding the first galaxies. We use
the characteristic radius $R$ in Sec.3.1 to denote this restriction,
i.e. the cross-correlation turns to be zero on scales smaller than $R$.  
In the calculation of the angular power spectrum, the power vanishes
if $\ell>2\pi r/R$.

\subsection{Cross-correlation: $C_{\rm em-ab}(\ell)$}

Before the EOR, most of IGM is still neutral, and the mean global 
fraction of ionized hydrogen at $z\approx20$ is well below unity. So, 
one may have an intuition that there should exist a strong
correlation between the 21cm emission generated by neutral hydrogen 
surrounding the ionized bubbles of first galaxies/quasars and 
the 21cm absorption from the uncollapsed, diffuse neutral hydrogen.
This arises because both the collapsed and uncollapsed baryons
trace the common large-scale density perturbations of the universe. 
Yet, the co-existence of 'hot' IGM around first galaxies and 'cold' IGM 
within the same region vanishes rapidly with the formation of more  
first generation stars, unless the inhomogeneous nature of Ly$\alpha$
and/or X-ray backgrounds are introduced (\citealt{barkana}; 
\citealt{pritchard07}; \citealt{santos};
\citealt{mesinger}). 
Recall that first galaxies formed in overdense regions, 
in which the heating of 
ambient diffuse IGM by energy release of the first stars 
proceeds more efficiently than that in underdense regions. 
In other words, the mean neutral fraction of the universe should not 
be used as a unique indicator of the significance of the cross correlation
$\langle T_{\rm em}T_{\rm ab}\rangle$. Indeed, the signature of 
cross-correlation $\langle T_{\rm em}T_{\rm ab}\rangle$  
on the total 21cm cosmic angular power spectrum could only be 
detected during the transition phase. This arises because the 
population of the first stars even in the overdense regions 
are rare during this rapid transition process
and most of the radiation energy is not yet delivered to the 
uncollapsed IGM in the same regions.
Such a cross-correlation makes a negative contribution
to, and may therefore suppress the total 21cm cosmic angular power 
spectrum. The cross-correlation can be represented by
%25
\begin{equation}
\langle T_{\rm em}T_{\rm ab}\rangle=T_0(1+\delta)x_{\rm H}
       \left(\frac{dT_{\rm ab}}{d\delta}\delta\right).
\end{equation}
The angular power spectrum weighted by window function $W_{\nu}(r)$ and
under small-angle approximation is
%26
\begin{equation}
C_{\rm em-ab}(\ell)= T_0 \int \frac{W_{\nu}^{2}(r)}{r^2} 
     \left(\frac{dT_{\rm ab}}{d\delta}\right)
     \left[P_{x_{\rm H}\delta}(\ell/r)+\bar{x}_{\rm H}f P_{\rm M}
     (\ell/r)\right]dr.
\end{equation}
Note that the ionized bubbles surrounding the first galaxies 
are excised in the above computation to account for 
the null effect of 21cm absorption from neutral hydrogen.

\subsection{Cross-correlation: $C_{\rm mh-ab}(\ell)$}

Unlike the above two cross-correlations which may manifest themselves
only in the rapid transition phase,   
the 21cm emission of minihalos should naturally exhibit a 
correlation with the 21cm absorption of uncollapsed IGM through the
entire period of the Dark Ages till the Grey Ages. 
The cross-correlation between these two components is
%27
\begin{equation}
\langle T_{\rm mh}T_{\rm ab}\rangle =
        \left(\frac{\lambda^3}{2k_{\rm B}}\right)
        \frac{(1+z)^2r^2}{H(z)}
        \left\{\int_{\rm Mmin}^{\rm Mmax}F\left
        [\frac{d^2n(M,z)}{dMdV}\right]dM\right\}
      \left(\frac{dT_{\rm ab}}{d\delta}\delta\right).
\end{equation}
The fluctuation arises purely from the density perturbation $\delta$, 
and therefore its power spectrum is simply
%28
\begin{equation}
P_{\rm mh-ab}({\mbox{\boldmath $k$}})=\left(\frac{\lambda^3}{2k_B}\right)
        \frac{(1+z)^2r^2}{H(z)}  
        \left(\frac{dT_{\rm ab}}{d\delta}\right)
        f P_{\rm M}({\mbox{\boldmath $k$}}).
\end{equation}
Finally, we can estimate the angular power spectrum modulated by
window function $W_{\nu}(r)$ through
%29
\begin{equation}
C_{\rm mh-ab}(\ell)=\left(\frac{\lambda^3}{2k_{\rm B}}\right)
        \int \frac{(1+z)^2W_{\nu}^{2}(r)}{H(z)} 
        \left(\frac{dT_{\rm ab}}{d\delta}\right)
        f P_{\rm M}(\ell/r) dr.
\end{equation}
One may argue that the fraction of neutral hydrogen in the above 
calculations may be overestimated because the contribution of 
minihalos to the total baryonic matter of the universe has not been 
excluded in the treatment of the uncollapsed IGM. To clarify the point, 
we have calculated the hydrogen mass inside minihalos at different
redshifts and found that its contribution to the total fraction of 
global neutral hydrogen is negligibly small.

%%%%%%%%%%%%%%%%%%%%%%%%%%%%%%%%%%%%%%%%%%%%%%%%%%%%%%%%%%%%%%%

\section{Results}

We adopt a frequency resolution of $\Delta\nu=0.1$MHz to conduct
our numerical computations of the angular power spectra of 
three auto-correlations and three cross-correlations. The total
21cm angular power spectrum is then obtained by summing up the
six components following Eq.(3).  We demonstrate all the resultant
angular power spectra in the conventional way,
$\delta T=[\ell(\ell+1)C(\ell)/2\pi]^{1/2}$, which provides a 
direct measurement of rms fluctuation of the 21cm brightness 
temperature in the corresponding multipole range.

\begin{figure}
   \begin{center}
   \psfig{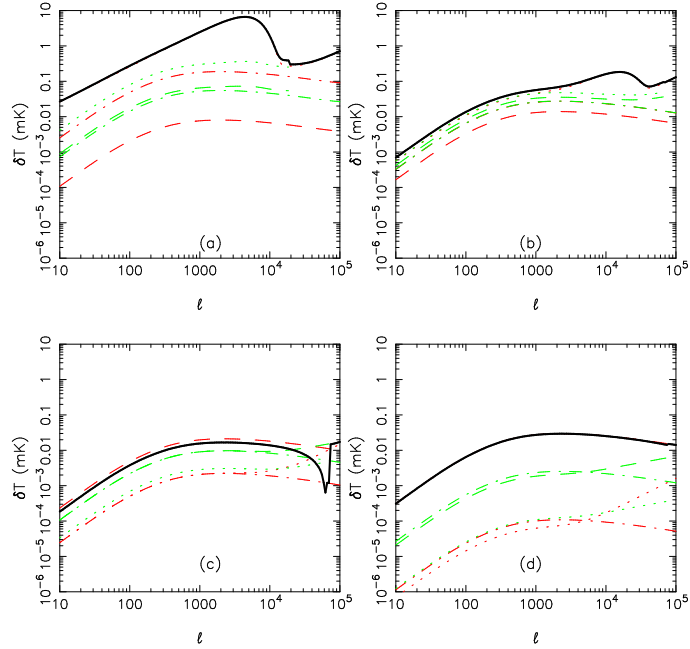}
   \end{center}
   \caption{The predicted angular power spectra of 21cm brightness fluctuations
   at four different redshifts, (a) $z=10$, (b) $z=15$, (c) $z=20$
   and (d) $z=25$. Three auto-correlations (red lines) and three 
    cross-correlations (green lines) are illustrated by 
   dotted curve ($C_{em-em}$), 
   dot-dashed curve ($C_{mh-mh}$), long dashed curve ($C_{ab-ab}$), 
   dotted curve ($C_{em-mh}$), dot-dashed curve 
   ($C_{mh-ab}$), and long dashed curve ($C_{em-ab}$), respectively. 
   Solid black curve represents the total angular power spectrum. 
   Note that each of the cross-correlation components $C(\ell)$ 
   has been multiplied by a factor of $2$ to account for its effective 
   contribution to the total power spectrum [see Eq.(3)]. 
   For the two negative cross-correlation components relevant 
   to the uncollapsed IGM, their absolute values are utilized.}
   \end{figure}

Figure 1 shows the  angular power spectra of three 
auto-correlations ($C_{\rm em-em}$, $C_{\rm mh-mh}$ and $C_{\rm ab-ab}$ )
and three cross-correlations  ($C_{\rm em-mh}$, $C_{\rm em-ab}$ and 
$C_{\rm mh-ab}$) calculated
at four different redshifts, z=10, 15, 20 and 25. The results of the three 
auto-correlations are certainly not novel and have already been 
demonstrated in the literature (see, for examples, 
\citealt{iliev02}; \citealt{loeb}; 
Zaldarriaga et al. 2004). 
As what is expected, the overall power spectrum for $\ell$ up to $10^5$ 
is governed by the auto-correlation term, $C_{\rm em-em}$ at lower 
redshifts and $C_{\rm ab-ab}$ at higher redshifts, respectively, 
and the transition occurs at $15<z<20$. The auto-correlation of 
minihalos contributes only a small fraction to the total power spectrum. 
Nonetheless, the cross-correlation of minihalos with other two mechanisms 
constitutes the second largest components to the total power 
spectrum: $C_{\rm em-mh}$ at redshifts below $z\approx 15$ and 
$C_{\rm mh-ab}$ at redshifts beyond $z\approx 20$,  though  
their absolute values are still small as compared with the contribution 
of corresponding auto-correlation ($C_{\rm em-em}$ at lower redshifts 
and $C_{\rm ab-ab}$ at higher redshifts).

\begin{figure}
\begin{center}
\psfig{file=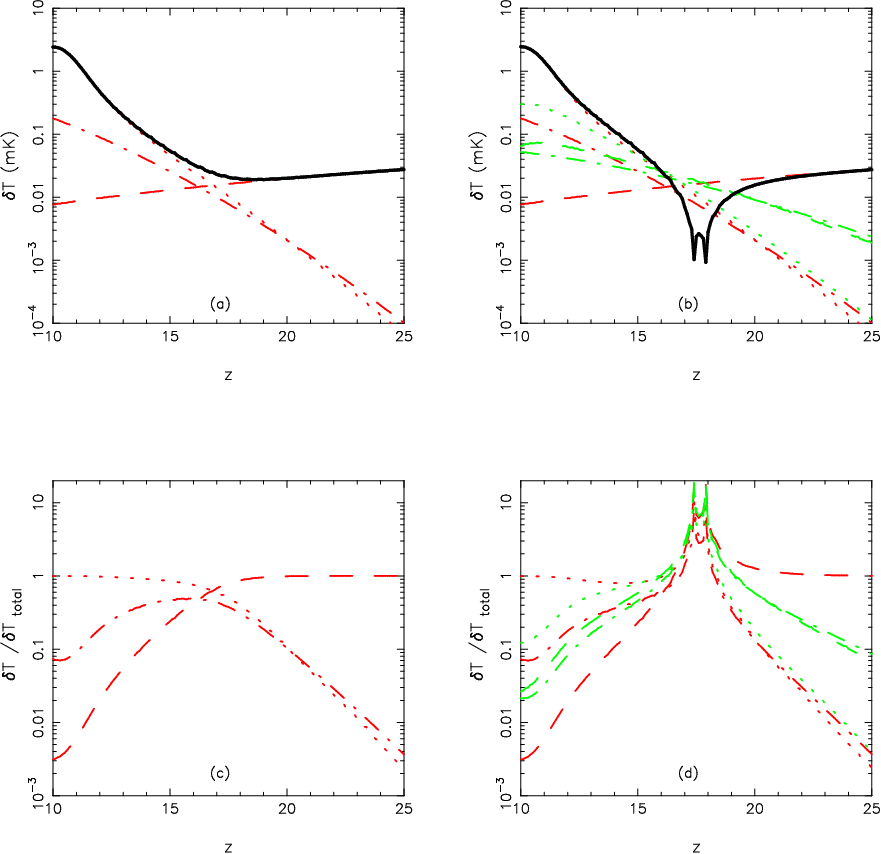,width=9.0cm}
\end{center}
\caption{ Redshift dependence of 21cm angular power spectra at 
multipole $\ell=10^3$. The left panels (a, c) and right 
panels (b, d) show the results without and with the inclusion 
of cross-correlations, respectively. The top panels (a, b) 
are the angular power spectra, while the bottom panels (c, d) 
display the fraction of each component in the total angular 
power spectra. We use the same notations for all the curves 
as in Fig.1.}
\begin{center}
\psfig{file=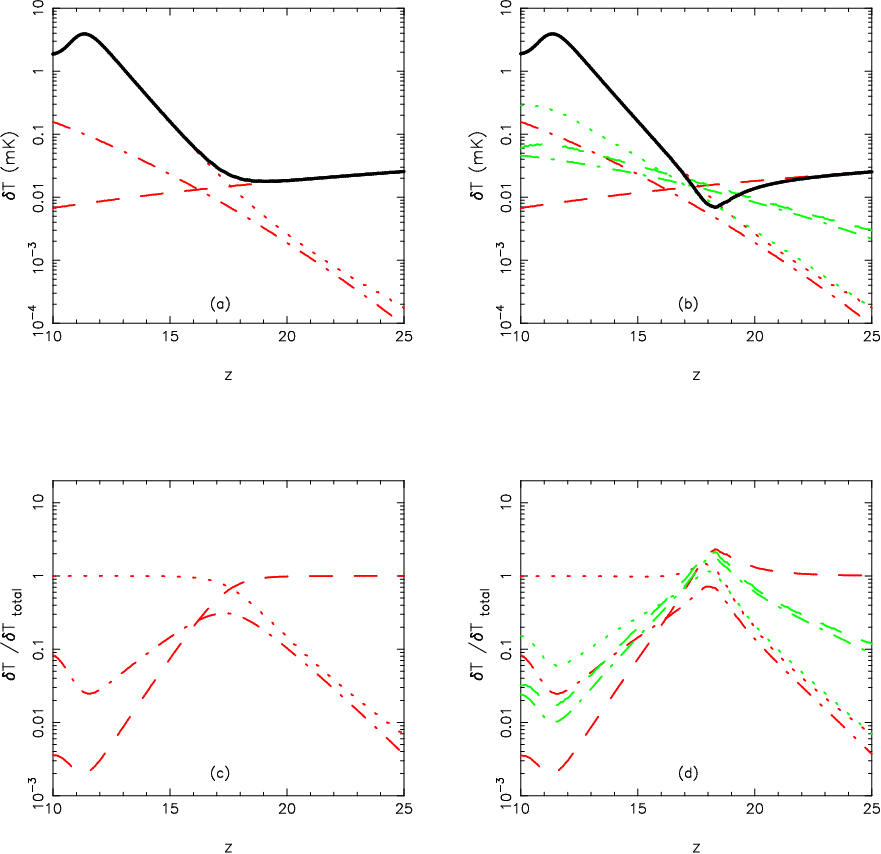,width=9.0cm}
\end{center}
\caption{ The same as Fig. 2 but for $\ell=10^4$. }
\end{figure}

While the effect of cross-correlations on the overall 21cm angular 
power spectrum is insignificant, the cross-correlation may exceed 
the auto-correlation during the transition phase which occurs 
at a certain redshift  between $z\approx15$ and $z\approx20$ 
in terms of the present models.  
To demonstrate the point clearly, we take the 21cm 
power spectrum of each component at two multipoles, $\ell=10^3$ and 
$\ell=10^4$, and illustrate in Figure 2 and Figure 3 
its time evolution as well as its fractions 
in the total power spectra. Note that under
certain circumstances some fractions may be larger than unity 
because of the negative contribution of the two terms,
$C_{\rm em-ab}$ and  $C_{\rm mh-ab}$, to the total power, which 
is actually a good indicator of the significance of cross-correlations at
the corresponding multipole range. For comparison, we have also displayed  
the results when only the auto-correlations are taken into account.
It turns out that inclusion of the cross-correlations leads to a 
suppression of the angular power spectrum during the transition phase, and
the extent and position of such a suppression varies with multipole $\ell$.
For the example of $\ell=10^3$ shown in Figure 2, there is a drop of
up to an order of magnitude around $z=17-18$ with regard to the 
angular power spectrum without the inclusion of
cross-correlations for the same $\ell$.  

\begin{figure}
\begin{center}
\psfig{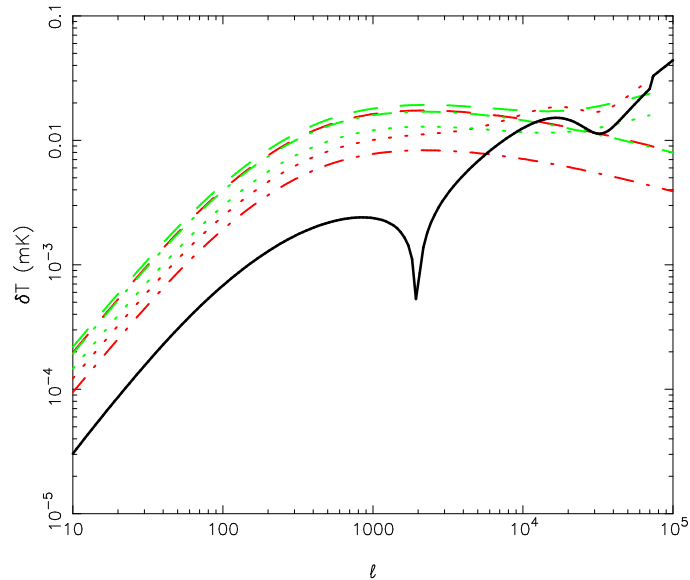}
\end{center}
\caption{Contribution of each correlation component to the 
total angular power spectrum at the transition phase of 
$z=17.5$. The same notations as in Fig.2 are used for 
all the curves. Note again that the absolute values of 
$C_{\rm em-ab}$ and  $C_{\rm mh-ab}$ are shown.}
\end{figure}

It is not difficult to understand the cause for the above power suppression
during the transition phase. If we neglect the contribution of minihalos, 
the total power spectrum reduces to
$C(\ell)=  C_{\rm em-em}(\ell)+C_{\rm ab-ab}(\ell)-2|C_{\rm em-ab}(\ell)|$.
Before and after the transition,  $C(\ell)$ is dominated by either
$C_{\rm em-em}(\ell)$ or $C_{\rm ab-ab}(\ell)$, depending on redshifts.
During the transition phase, however, the two effects make more or 
less the same contribution to the total power spectrum, or 
the auto-correlation is roughly the same as the cross-correlation 
in amplitude, hence reducing greatly the power. In other words,  
the two components almost cancel each other because the cross-correlation
has a negative sign.  We illustrate the point in Figure 4 by plotting
the angular power spectrum at redshift $z=17.5$, the transition phase
in terms of our present models.  It appears that a combination of 
the auto- and cross-correlations yields a much lower angular power spectrum.

%%%%%%%%%%%%%%%%%%%%%%%%%%%%%%%%%%%%%%%%%%%%%%%%%%%%%%%%%%%%%%%

\section{Discussion and Conclusions}

Using a simple analytic prescription of both 21cm emission and absorption of
neutral hydrogen in the epoch of the Grey Ages, we have calculated
the cross-correlation between three major mechanisms of cosmic 21 cm 
radiation, namely, the emission from neutral hydrogen surrounding 
the ionized bubbles of first galaxies, the emission of neutral hydrogen
inside minihalos, and the absorption of diffuse neutral hydrogen 
against the CMB radiation. It has been shown that inclusion of the three
cross correlations does not alter the overall angular power spectrum of
cosmic 21cm background. However, during the transition 
of 21cm radiation field from emission-dominated phase to 
absorption-dominated phase, the cross-correlation may play a key role 
in the determination of the shape and magnitude of the 21cm power
spectrum. This arises primarily from the negative contribution of
cross-correlation between the absorption of diffuse IGM and the other 
two emission mechanisms, which leads to a significant power suppression
on the fluctuation of 21cm radiation field during
the transition phase. In our naive models for the above three 
mechanisms, especially the oversimplified scenario for the 21cm 
emission from the IGM associated with first galaxies, the transition
occurs around redshift $z\approx17-18$, which varies slightly with
multipole $\ell$.  

The effect of cross-correlations on the 21cm angular power spectrum 
could be detected as a remarkable drop of power at a certain frequency
range corresponding to the transition phase. The precise value of
the transition frequency depends on the combined effects of various 
mechanisms of 21cm radiation in the epoch of the Grey Ages.
An extension of the present work to the adoption of more sophisticated
models for the 21cm radiation field at the Grey Ages may provide a 
more accurate prediction of the 21cm angular power spectrum.  

Finally, we must caution that the effect of the cross-correlations 
on the total cosmic 21cm power spectrum could be considerably
exaggerated as a result of our 'toy model', and some of 
our numerical results may even be artificial because of the
'artificial' split between the three major mechanisms of 21cm radiation.  
Numerical simulation should be applied to clarify the issue. 

%%%%%%%%%%%%%%%%%%%%%%%%%%%%%%%%%%%%%%%%%%%%%%%%%%%%%%%%%%%%

\section*{Acknowledgments}
We gratefully acknowledge an anonymous referee for valuable 
comments. This work was supported by the Ministry of Science and 
Technology of China, under Grant No. 2009CB824900.

%%%%%%%%%%%%%%%%%%%%%%%%%%%%%%%%%%%%%%%%%%%%%%%%%%%%%%%%%%%%%%%%%%%%%%%%%%%%%%%%%%%%%%%%%%%%%

%%%%%%%%%%%%%%%%%%%%%%%%%%%%%%%%%%%%%%%%%%%%%%%%%%%%%%%%%%%%%%%%%%%%%%%%%%%%%%

\label{lastpage}
\end{document}